# Time-resolved second harmonic generation study of buried semiconductor heterointerfaces using soliton-induced transparency

Y. D. Glinka,[1,2*] N. H. Tolk,[3] J. K. Furdyna[4]

[1]*Institute of Physics, National Academy of Sciences of Ukraine, Kiev, 03028, Ukraine*
[2]*Department of Physics, University of Texas at Austin, Austin, Texas 78712, USA*
[3]*Department of Physics and Astronomy, Vanderbilt University, Nashville, Tennessee 37235, USA*
[4]*Department of Physics, University of Notre Dame, Notre Dame, Indiana 46556, USA*

The transient second harmonic generation and linear optical reflectivity signals measured simultaneously in reflection from GaAs/GaSb/InAs and GaAs/GaSb heterostructures revealed a new mechanism for creating self-induced transparency in narrow bandgap semiconductors at low temperatures, which is based on the dual-frequency electro-optic soliton propagation. This allows the ultrafast carrier dynamics at buried semiconductor heterointerfaces to be studied.

Once the first nonlinear optics experiment on second harmonic generation (SHG) in a quartz crystal has been realized [1], the effect has quickly received wide recognition from the semiconductor community since can provide a non-destructive method of surface structural analysis. The basic idea behind such a SHG application has been proposed by Bloembergen *et al.* [2], who suggested that SHG measured in reflection from the semiconductor surface in comparison to the usual linear optical reflectivity (LOR) is governed by a tensor quantity that contains elements of the bulk crystal symmetry. The spatial (surface) sensitivity of the SHG response from centrosymmetric semiconductors, such as Si and Ge, is due to that the lowest-order contribution (electric-dipole) is symmetry allowed only for a near-surface layer. Alternatively, SHG in reflection from non-centrosymmetric semiconductors, such as GaAs and GaSb, is electric-dipole allowed for both surface and bulk contributions. However, despite the absorption length in GaAs and GaSb at the fundamental frequency ($\omega_1$) of Ti:Sapphire laser usually used for measurements is around 1000 and 200 nm, the SHG photon absorption at $\omega_2 = 2\omega_1$ is much larger and hence limits the photon escape depth to only a few tens of nanometers (50 and 15 nm), respectively. This still holds the spatial sensitivity of SHG for non-centrosymmetric semiconductors whereas restricts it to be applied to study the buried heterointerfaces far distanced from the sample surface.

Nevertheless, we have recently demonstrated that the transient SHG and LOR responses from GaAs/GaSb/InAs heterostructure at low temperatures follow well the fundamental absorption edge of GaAs despite the buried heterointerface is created with thick (500 nm) GaSb [3-5]. The transient responses are induced by the interfacial electric field arising at the buried heterointerfaces owing to the charge separation. As a result, the electric-field-induced SHG and both linear and quadratic electro-optic effects allow for monitoring the ultrafast carrier dynamics at the buried heterointerfaces. However, the mechanism for creating transparency in GaSb that would allow the laser excitation to be delivered to the buried heterointerfaces and the SHG and LOR photons to be escaped from them still remains unknown.

In this Letter we report the experimental results on time-resolved SHG and LOR measured in reflection from the MBE grown GaAs(100nm)/GaSb(500nm)/InAs(20nm) and GaAs(100nm)/GaSb(400nm) heterostructures, which point to a new type of self-induced transparency in highly absorbing narrow bandgap semiconductors occurring at low temperature due to the quasi-ballistic transport of photoexcited carriers. The mechanism takes account of the photo-Dember field solitary wave, which traps both the fundamental and SHG pulses, slowing their velocity down to that of the solitary wave. The trapped light pulses maintain the amplitude of the solitary wave and hence create a condition, at which the self-reinforcing nonlinear optical polarization (dual-frequency electro-optic soliton) can freely propagate through the semiconductor. Among well-known mechanisms for creating self-induced transparency, the mechanism discussed here is most closely related to that of based on photorefractive soliton propagation in the biased photorefractive medium [6].

We used the combined ultrafast technique implying the simultaneous measurements of transient SHG and LOR in pump-probe configuration, which has initially been proposed to study the photoexcited plasma-induced non-thermal structural phase transition in semiconductors [7,8] and then extended to study the ultrafast carrier dynamics at buried heterointerfaces [5]. The initial beam of a mode-locked Ti:Sapphire laser of 150-fs pulses tunable in the range of 1.42 - 1.55 eV was split into two cross polarized beams, which were focused into ~ 100-μm spot on the sample surface. The pump beam with power $P = 20 – 188$ mW (after chopping at a frequency of 800 Hz) was incident normally ($z$ axis), whereas the $p$ polarized probe beam of 80 mW power was optically delayed and directed to the sample surface at $\theta_1 = 45°$ [Fig. 1(a)]. According to Snell's law the incident angle in the heterostructure is reduced to $\theta_2 = 9°$. The optically separated SHG and LOR responses were detected with pump-to-probe delay-time by a photodiode and a photomultiplier tube, respectively, using the phase-matched "lock-in" amplifiers.

Fig. 1 (b) and (c) shows the transient signals measured at 4.3 K with photon energies near the GaAs bandgap ($E_g = 1.52$ eV). The transient SHG signal from GaAs/GaSb/InAs

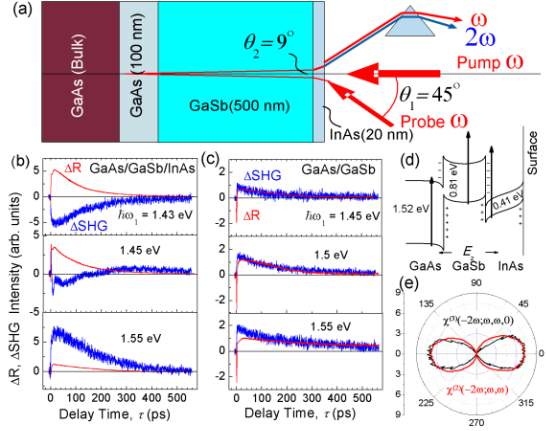
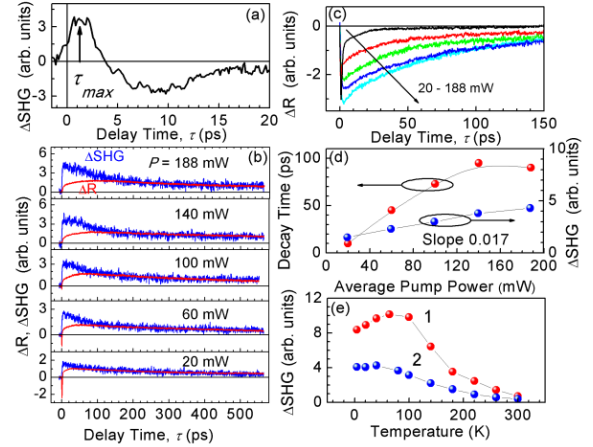

FIG. 1 (color online). (a) The depth section of the heterostructure and the experimental arrangement used. (b) and (c) Transient SHG and LOR signals from GaAs/GaSb/InAs and GaAs/GaSb heterostructures measured in $p_{in}/p_{out}$ polarization geometry at 4.3 K with photon energy indicated and $P$ = 130 and 20 mW, respectively. (d) Band alignment of GaAs/GaSb/InAs heterostructure. (e) The normalized SHG incident polarization angle rotational pattern for GaAs/GaSb heterostructure measured at 4.3 K with $\hbar\omega_1$ = 1.55 eV for the time-independent (background) SHG [$\chi^{(2)}(-2\omega;\omega,\omega)$] and transient SHG signal taken with $P$ = 188 mW at delay-time of 6 ps [$\chi^{(3)}(-2\omega;\omega,\omega,0)$].

FIG. 2 (color online). (a) The short-delay-time part of the transient SHG signal from Fig. 1 (b) taken with $\hbar\omega_1$ = 1.45 eV. (b) Transient SHG and LOR signals from GaAs/GaSb heterostructure measured in $p_{in}/p_{out}$ polarization geometry at 4.3 K with $\hbar\omega_1$ = 1.55 eV and $P$ indicated. (c) The absorption bleaching component of the transient LOR signal extracted from traces shown in (b). (d) The pump power dependence of the absorption bleaching decay-time and the transient SHG signal from GaAs/GaSb heterostructure. (e) Temperature dependence of the transient SHG signal from GaAs/GaSb/InAs (1) and GaAs/GaSb (2) heterostructures measured with $\hbar\omega_1$ = 1.55 eV and $P$ = 188 mW.

heterostructure reveals a signal sign flip with increasing photon energy whereas the corresponding transient LOR signal only gradually weakens. The signal trends are well consistent with our previous measurements [3-5]. Because of two heterointerfaces in GaAs/GaSb/InAs heterostructure with the oppositely directed transient interfacial electric fields, $\vec{E}_{z1}$ and $\vec{E}_{z2}$ [Fig. 1(d)], the total signal results from the interference of signals induced at both heterointerfaces [5]: $\Delta SHG = -6B\left\{\left[\Delta^{(2)}_{zzz}\chi^{(3)}_{zzzz}\vec{E}_{z1}\right] - \left[\Delta^{(2)}_{zzz}\chi^{(3)}_{zzzz}\vec{E}_{z2}\right]\cos\phi\right\}(I^\omega)^2$, where $B$ is a constant, $\Delta^{(2)}_{zzz}(-2\omega;\omega,\omega)$, and $\chi^{(3)}_{zzzz}(-2\omega;\omega,\omega,0)$ are the second- and third-order heterointerface susceptibilities, all refer to GaSb since the out-of-resonance value of the third-order susceptibility for GaSb is at least one order of magnitude larger than that for InAs and GaAs [9]; $\phi$ is the relative phase difference between the contributions: $\phi = \left[(2\pi/\lambda_1)n_1 L/\cos\theta_2\right] + \left[(2\pi/\lambda_2)n_2 L/\cos\theta_2\right] = 2\pi \cdot 6.867$, where $\lambda_1$ and $\lambda_2$ are the light wavelength of 867 nm (1.43 eV) and 433.5 nm (2.86 eV), $n_1$ and $n_2$ are the corresponding refractive indices of GaSb, $L$ is the distance between the heterointerfaces (500 nm). Since $\phi$ is close to integer multiple of $2\pi$, it does not affect significantly the signal sign correlation. We consider only the isotropic tensor components since the isotropic contribution to the depletion electric field induced SHG from GaAs(001) surface has been evidenced to be dominant over the others [10]. This is also confirmed by the incident polarization angle rotational pattern [Fig. 1(e)], which reveals the simplest two-fold symmetry corresponding to the $p_{in}/p_{out}$ polarization geometry since the electric field of only $p$ polarized light wave give projections onto the surface normal. The photon energy dependence points hence to whether or not the electric field at the GaAs/GaSb heterointerface contributes into the total signal. Because of the monocomponent nature of the transient SHG signal, it is fully reversible with increasing photon energy. The characteristic ultrafast spike at $\tau_{max}$ = 1.27 ps induced with $\hbar\omega_1$ = 1.45 eV [Figs. 1(b) and 2(a)] indicates the interplay of the two contributions and allows to estimate the time required for the laser pulse to propagate through the 500-nm thick GaSb: $\tau_P = (|\tau_{R2} - \tau_{R1}|/2\tau_{R1})\tau_{max}$, where $\tau_{R1}$ = 2.93 ps and $\tau_{R2}$ = 0.71 ps are the rise-time constants obtained from the fit of the individual signals from GaSb/InAs and GaAs/GaSb heterointerfaces taken with $\hbar\omega_1$ = 1.43 and 1.55 eV, respectively. The obtained value of $\tau_P$ = 0.48 ps gives the velocity of the laser pulse propagation of $\upsilon_P = L/\tau_P = 1.04\times10^6$ m/s, which well matches the estimate of the electron quasi-ballistic velocity in polar semiconductors [11], whereas is at least one order of magnitude less than the group velocity of light in GaSb: $\upsilon_G$ = 4.8×10$^7$ m/s (867 nm). This suggests that the laser pulse propagation in GaSb is controlled by the electronic excitation propagating in the same direction.

The transient LOR signal is multicomponental since additionally to the usual absorption bleaching component both the linear and quadratic electro-optic effects give rise into the total transient signal [5]:

$$\Delta R = 2C\left[\left(r_{zz}\vec{E}_{z1} + p_{zzzz}\vec{E}_{z1}\vec{E}_{z1}\right) - \left(r_{zz}\vec{E}_{z2} + p_{zzzz}\vec{E}_{z2}\vec{E}_{z2}\right)\cos\phi\right],$$

where $C$ is a constant, $r_{zz}(-\omega;\omega,0)$ and $p_{zzzz}(-\omega;\omega,0,0)$ are the effective electro-optic coefficients. The relative phase difference, $\phi = \left[(2\pi/\lambda_1)2n_1 d/\cos\theta_2\right] = 2\pi\cdot4.991$, can also be ignored. The opposite sign of the transient SHG and LOR signals originates from the relation between the nonlinear susceptibilities and electro-optic coefficients: $\Delta^{(2)}_{zzz} = -\varepsilon_{ii}\varepsilon_{jj}r_{zzz}/4\pi$ ; $\chi^{(3)}_{zzzz} = -\varepsilon_{ii}\varepsilon_{jj}p_{zzzz}/4\pi$ [5,12]. However, the existence of heterointerfaces introduces an additional mirror symmetry operation: $(x, y, z) \rightarrow (x, y, -z)$. This is equivalent to the situation, at which all the tensors reverse their signs with electric field reversal [5,12]. Specifically, the linear-in-field term will be always positive since $\vec{E}_z$ and $r_{zzz}$ flip their signs jointly. For the same reason, both the quadratic-in-field part of the transient LOR signal and the transient SHG signal are field-direction-dependent due to the product of an odd number of matrices. As a result, the sign of the transient SHG response is opposite to only that of the quadratic-in-field part of the transient LOR response. This explains why the transient LOR signal is not fully reversible with increasing photon energy [Fig. 1 (b)].

Since there is only one heterointerface in GaAs/GaSb heterostructure, the amplitude of the SHG and LOR signals simply follows the absorption edge of GaAs [Fig. 1 (c)]. The transient SHG signal from GaAs/GaSb heterostructure is at least three-fold weaker than that from the GaAs/GaSb/InAs one. This suggests that the linear electro-optic effect mainly contributes into the transient LOR signals. The absorption bleaching component, being always negative, shows a short decay-time constant of ~ 2 ps. The latter rises up with increasing photon energy causing an increased rise-time constant of the total LOR signal since the interplay with the positive electro-optic contribution. This appears as a difference in shape between the transient SHG and LOR signals with increasing photon energy [Fig. 1 (c)] and allows the absorption bleaching component to be extracted from the total LOR signal. The short decay-time of the absorption bleaching component is well consistent with the lifetime of free carriers in the highly excited band states of GaSb and InAs, which is mainly controlled by the longitudinal optical (LO) phonon emission [13]. Since the time required to emit a single LO-phonon by hot electrons in GaSb ($\tau_{LO}$ ~ 450 fs) is three times longer than the excitation pulse [4], the initial electron power loss, $P_e \approx (\hbar\omega_{LO}/\tau_{LO})\exp(-\hbar\omega_{LO}/k_B T_e)$, where $T_e$ is the electron temperature [13], is negligibly small. The power loss due to the elastic hot electron scattering with acoustic phonons is negligible too for the energy difference reason. This creates a condition for the quasi-ballistic electron transport, which prevents against the carrier spatial buildup and hence reduces their Drude absorption efficiency. The free-carrier absorption (Drude absorption) of density $N$ in GaSb at room temperature can be estimated as $\alpha_D$ (cm$^{-1}$) $\approx 2\times10^{-17} N$ (cm$^{-3}$) [14]. The maximal power used in our measurements generates the carrier density of $N = 5.4\times10^{17}$ cm$^{-3}$ and hence the suppression of the carrier quasi-ballistic transport leads to a non-recoverable damage of the sample surface observed at room temperature.

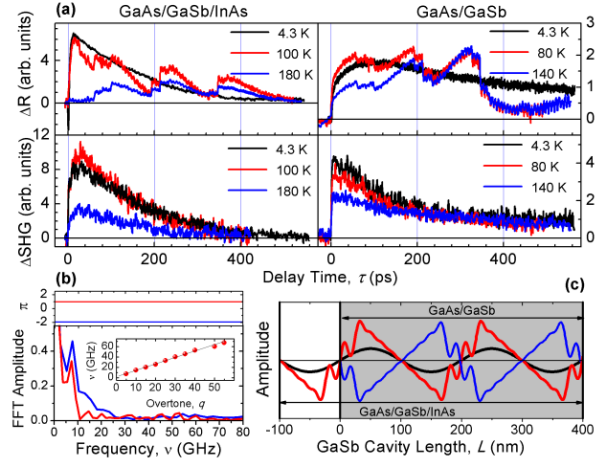

FIG. 3 (color online). (a) Time-resolved LOR and SHG signals for GaAs/GaSb/InAs and GaAs/GaSb heterostructures as a function of temperature indicated. (b) Fourier analysis of LOR responses for GaAs/GaSb/InAs (red) and GaAs/GaSb (blue) heterostructures. Insert shows the overtone frequencies. (c) The result of numerical modeling of the standing acoustic waves in GaAs/GaSb/InAs and GaAs/GaSb heterostructures, which takes into account the acoustic wave overtones shown in (b). The single mode standing wave is shown in black.

The transient signals rise up with the pump power but with very little slope [Fig. 2 (b) and (d)]. This suggests that the nonlinearity saturation regime is reached. Also, the absorption bleaching contribution becomes more prominent [Fig. 2 (b) and (c)] owing to a significant increase in the decay-time constant, which finally reaches its saturation point as well [Fig. 2 (d)]. The latter is a signature of the hot LO phonon bottleneck occuring at low temperature [15], which is controlled by the relaxation of LO phonons through the anharmonic three phonon (Klemens/Ridley) decay process involving acoustic phonon branches [16]. The phonon bottleneck diminishes with increasing temperature owing to activation of the Klemens/Ridley decay channel. The latter process manifests itself in the formation of the standing acoustic waves in the cavity created within the GaSb layer, the contribution of which is superimposed on the total LOR response through the Brillouin scattering mechanism. This leads to the characteristic sawtooth waveform to appear in the transient LOR signals with increasing temperature [Fig. 3 (a)], which points to the multimode nature of the standing wave formed [17]. Here we note that the standing acoustic wave appears as an individual amplitude component since the phase-sensitive detection is used. Alternatively, the temperature rise does not affect the profile of the transient SHG signal [Fig. 3 (a)], suggesting the absence of parametric interaction between $\omega_1$ and $\omega_2$ waves [18]. The temperature dependence of the transient SHG signal intensity shows a smooth bell-like shape [Fig. 2 (e)], which is typical for the carrier mobility in polar semiconductors [19] and hence supports the quasi-ballistic regime of the carrier transport at low temperatures.

One can see that an elongation of the GaSb cavity by 100 nm in GaAs/GaSb/InAs heterostructure changes the phase of the sawtooth waveform by $\pi$. This is consistent with Fourier analysis shown in Fig. 3 (b) and allows for numerical modeling of the standing acoustic wave [Fig. 3 (c)]. Specifically, the dominant peaks of frequencies $\nu$ = 7.77 (GaAs/GaSb) and 7.33 GHz (GaAs/GaSb/InAs) correspond to the 4 and 5 harmonic of the fundamental frequency of the acoustic wave, respectively. The much weaker higher frequency modes are the overtones ($q$) of the dominant acoustic mode. The shape of the standing wave depends on both the amplitude and the number of overtones involved [17]:
$$\Delta R(\tau) = \sum_q (1/q)\cos\{2\pi q\nu\tau + [(50+L)\pi/100] + q\varphi_d\},$$
where the phase $\varphi_d$ = 0, $\pm\pi/4$ refers to the diffraction of the longitudinal acoustic waves generated [20], three values of which causes the fine structure of the sawtooth standing waveforms [Fig. 3 (a) and (c)].

The sound velocity ($c_a$) then can be estimated from the cavity decay time, which for GaAs/GaSb heterostructure, for example, is as follows: $\tau_c = (1/\nu) = 2L/4c_a\Lambda$, where $\Lambda$ is the cavity intensity losses, which similarly to the optical resonators can be taken as $\Lambda = 2 - R_1 - R_2$, where $R_1$ = 1 is the sound reflectivity coefficient for the GaSb/vacuum interface and $R_2$ is that for GaAs/GaSb interface: $R_2 = (z_2 - z_1)/(z_2 + z_1)$ = 0.065, where acoustic impedances for GaAs ($z_2$ = 25.4×10$^6$ kg/s·m$^2$) and GaSb ($z_1$ = 22.3×10$^6$ kg/s·m$^2$) have been taken into account. The resulting losses and sound velocity then are: $\Lambda$ = 0.935 and $c_a$ = 1.66×10$^3$ m/s. The latter value is about half the velocity of longitudinal acoustic wave in GaSb (3.97×10$^3$ m/s) and closely matches that for a shear (transverse) wave: $c_A = \sqrt{G/\rho}$ = 1.79×10$^3$ m/s, where $G$ is shear modulus and $\rho$ is density of GaSb [21]. The shear wave observation unambiguously proves the diffraction of the initially generated longitudinal acoustic waves [22] and an extended source of their generation, which we associate with the trace of the electronic excitation propagating inward the medium with quasi-ballistic velocity. It is clear that the local generation of acoustic waves at the surface cannot explain the ultrafast carrier dynamics occurring in the heterostructures since the sound velocity in the materials is too low.

Combining all these facts together, the mechanism for creating transparency in GaSb can be saggested as that being related to the ultrafast buildup of the photo-Dember electric field, $\vec{E}_D$ [23], which is known to be enhanced in narrow bandgap semiconductors at low temperatures owing to high carrier mobility. In the dissipating media the photo-Dember field is modeled as a decaying solitary wave propagating inward the semiconductor from the surface with an average velocity of ~ 1.0×10$^6$ m/s [24]. As a result, in the strongly absorbing medium the laser pulse propagation is exclusively governed by the photo-Dember field solitary wave, which leads to the local nonlinear refractive-index variation ($\delta n_1 = 0.5 n_1^3 r_{zzz} \vec{E}_D$) and consequently to the laser beam self-focusing. The latter process balances the natural beam diffraction and hence creates the transient self-reinforcing nonlinear optical polarization (electro-optic soliton). This can be imagined as a trap of the laser pulse at frequency $\omega_1$ by the photo-Dember field solitary wave, which is maintained by the laser pulse propagation and causes slowing speed of light down to that of the solitary wave.

The simultaneously generated SHG wave at frequency $\omega_2$ is trapped by the solitary wave too through the third-order nonlinearity and travels about the same direction, however, without optical parametric interaction with the wave at frequency $\omega_1$ due to their wave vector mismatch. The electro-optic soliton is hence a dual-frequency soliton, in which both $\omega_1$ and $\omega_2$ beams propagate with the same velocity equal to the carrier ballistic velocity. The soliton can be reflected from the heterointerfaces being enhanced by the transient interfacial electric fields. This allows both the LOR and SHG photons to be escaped from buried heterointerfaces and measured in the time-resolved mode.

In conclusion, we have proven experimentally the existence of self-induced transparency in narrow bandgap semiconductors at low temperature, which is based on the dual-frequency electro-optic soliton propagation. This opens the way to a new opportunity for the time-resolved SHG studies of buried semiconductor heterointerfaces.